\documentclass{article}
\usepackage[english]{babel}
\usepackage{epsfig}
\graphicspath{{./Figures/}}
\usepackage[utf8]{inputenc}
\usepackage{floatrow}
\usepackage{amssymb,amsmath,amsfonts,amsthm,graphicx,psfrag}
\usepackage{indentfirst}
\usepackage{hyperref}

\usepackage[title,titletoc]{appendix}
\usepackage{graphicx}
\usepackage{amsfonts}
\usepackage{bm}
\usepackage{verbatim}
\usepackage{epsfig}

\graphicspath{{./pictures/}}

\usepackage{mathrsfs}
\usepackage{amsthm}
\usepackage{amssymb}
\usepackage[T1]{fontenc}

\usepackage{authblk}
\usepackage{cite}
\usepackage{slashed}
\usepackage{amsmath}
\usepackage[left=2cm,right=2cm,
    top=2cm,bottom=2cm,bindingoffset=0cm]{geometry}
\newcommand{\be}{\begin{eqnarray}}
\newcommand{\ee}{\end{eqnarray}}
\usepackage{authblk}

\title{The speed of sound in QGP and SU(3) Yang-Mills theory}
\author[*,+]{Z.V Khaidukov}
\author[*,+]{M.S Lukashov }
\author[*]{Yu.A.Simonov}

\affil[*]{Institute for Theoretical and Experimental Physics, B. Cheremushkinskaya 25, Moscow, 117259, Russia}
\affil[+]{Moscow Institute of Physics and Technology,
Institutskiy per. 9, 141700 Dolgoprudny, Moscow Region, Russia}

\begin{document}
  \maketitle
\begin{abstract}
The speed of sound $C_{s}$ in the SU(3) and (2+1)QCD  is calculated  within the Field Correlator Method using the nonperturbative colour magnetic confinement and Polyakov loop interaction in the deconfined region.The resulting  $C_{s}$ displays a discontinuity at $T=T_{c}$ in the SU(3) case. It is shown numerically and analytically that $ C^{2}_{s}$ never exceeds $\frac{1}{3}$ both for SU(3) and (2+1) QCD for vanishing chemical potential.A good agreement is found of our numerical results with the corresponding lattice data.
\end{abstract}

\section{Introduction}
One of the most significant discoveries of recent times was experimental detection of quark-gluon plasma(QGP) \cite{QGP1,QGP2,QGP3,QGP4}. It is the state of matter existing at extremely high temperatures in QCD. These conditions can be created in relativistic heavy-ion collisions\cite{QGP5,HIC7,HIC8,HIC9,HIC10}.

In the framework of the research, it has been found that QGP, contrary to early theoretical predictions\cite{HIC1,HIC2,HIC3,HIC4,HIC5,HIC6}, is a liquid with an extremely low viscosity, but  not a gas consisting of quarks and gluons. Therefore to describe processes in the heavy ions collision, (see for example \cite{Rev} and references therein.) it is possible to use a hydrodynamic approach  related to the physics of QGP \cite{Hydro1,Hydro2}. One of the most important characteristics in hydrodynamics is the value of the speed of sound in the medium and its dependence on parameters (temperature, density,etc.).For example, it governs the evolution of the fire-ball produced in the heavy-ion collision, and one of the most important observables for describing of QGP formation -the elliptic flow  \cite{Elle1,Elle2,Elle3,Elle4}. \par
The speed of sound is connected with the  conformal symmetry breaking in hot QCD. In a scale-invariant system in  the case of 3 spatial dimensions it should be $C^{2}_{s}=\frac{1}{3}$ , because the trace of momentum-energy tensor must vanish $\epsilon-3P=0$.
It  also can  carry some information about the type of phase transition in the system.\par
The main source of information on  the speed of sound in QCD is related with calculations on the lattice\cite{latt1,latt2,latt3,latt4}. In the confined phase  one can also use the hadron resonance gas model (HRG) \cite{HRG1,latt3,latticehadr1}. The independent way to obtain predictions for the speed of sound in QCD is connected with holographic description\cite{Holo1,Holo2,Holo3,Holo4,Holo5,Holo6,Holo7,Holo8}.\par
As it has become clear from numerous lattice and experimental studies of QCD at $T>0$, the main dynamics of  both hadron and  QGP phases  is of the nonperturbative(np) origin and should be treated within the np methods.The fundamental approach to the np QCD is developed in the framework of the Field Correlator Method(FCM) \cite{VCM2,VCM3,VCM4,VCM5,VCM6,VCM7} generalized to non-zero temperatures \cite{TVCM1,TVCM2,TVCM3,TVCM4,TVCM8}.
In this paper we use the FCM to calculate the speed velocity in case of finite temperatures and taking into account the colour magnetic confinement(CMC) as it was done in \cite{VCM1,TVCM5,TVCM6,TVCM7,TVCM9}. The strength of this method is the ability  in a self consistent manner to calculate the speed of sound both in the confinement phase and at temperatures above $T_C$ both at zero and non-zero chemical potential. The second case  is very important  because  of the ``sign problem'' in lattice calculations in this domain. To circumvent this difficulty  in the case of  $N_{c}=3$ in  QCD  one  finds  Taylor coefficients in expansion  around zero chemical potential   to obtain the information  about small densities\cite{thLatt1,thLatt2}, or uses imaginary  chemical potential\cite{thLatt3}, or else considers  the number of colours  $N_{c}=2$, where this problem  is absent\cite{Kot1,Kot2,Kot3,Kot4}. In this work we will calculate the  speed of sound  in the case of pure SU(3) gluodynamics and also in the presence of quarks in (2+1) QCD, for $\mu=0$

 \par The paper is organized as follows.  In section 2 we introduce the FCM in case of finite temperatures. In sections 3 and 4  we  use it  to define  the speed of sound as a  function of the temperature  in the case of pure Yang-Mills and of (2+1) flavors  QCD. Finally we summarize and discuss  the obtained results  in Section 5 .\par

\section{The Field Correlator Method}
The FCM  is a useful instrument to treat the physics outside the area of perturbative theory.   Analysis of  physics of QGP  in terms of FCM   made in \cite{TVCM1,TVCM2,TVCM3,TVCM4,TVCM8}, has shown the important role of Polyakov loops for description of thermodynamic of QGP.Below we also take into account CMC effects, which are especially important at high T.  The main idea is as follows:
 The gluonic field $A_{\mu}$ splits into the background field $B_{\mu}$ and the (valence
gluon) quantum field $a_{\mu}$: $A_{\mu}=B_{\mu}+a_{\mu}$, both satisfying the periodic boundary
conditions.
The partition function is:
\begin{eqnarray}
Z(B,T)=N\int{D\phi exp(-\int^{\beta}_{0} dt\int{ d^{3}x \mathcal {L}_{tot}})},
\end{eqnarray}
where $\phi$ denotes all set of fields $a_{\mu},\psi,\psi^{+}$ and ghost fields. In the lowest order
in $ga_{\mu}$ one may obtain a result in the so-called Single Line Approximation (SLA) \cite{TVCM4,TVCM8},where $q\bar{q}$ and gg correlations are neglected
\begin{eqnarray}
Z(B,T)=N_{1}[det(G^{-1})]^{-\frac{1}{2}}det(-D^{2}_{\lambda}(B))[det(m^{2}_{q}-\hat{D}^{2}(B))]^{1/2} ,
\end{eqnarray}
 where $N_{1}$-normalization factor, $D_{\lambda}(B)=\partial_{\lambda}-igB_{\lambda},G^{-1}=D^{2}_{\lambda}\delta_{\mu\nu}+2igF_{\mu\nu}$. 
The thermodynamic potential F(T) is connected to Z(B,T) in  standard
way
 \begin{eqnarray}
F(T)=-T\ln(Z(B))_{B}
\end{eqnarray}
where index B means   averaging over all background fields.
In SLA   the contributions of gluons and quarks  in F(T) are   separated:$$F(T)_{SLA} = F_{q}(T) + F_{gl}(T)$$.  

In FCM the  breaking of Lorentz invariance at finite temperatures becomes apparent through existence of two types of string tension:
 
\begin{eqnarray}
\sigma^{E,H}=\frac{1}{2}\int{D^{E,H}d^{2}z}
\end{eqnarray}
where $D^{E,H}$  is obtained from
\begin{eqnarray}
\frac{g^{2}}{N_{c}}\ll Tr E_{i}(x)\Phi E_{j}(y)\Phi^{+} \gg=\delta_{ij}(D^{E}(u)+D^{E}_{1}(u)+u^{2}_{4}\frac{\partial D^{E}_{1}(u)}{\partial u^{2}})+u_{i}u_{j}\frac{\partial^{2} D^{E}_{1}(u)}{\partial u^{2}} \\
\frac{g^{2}}{N_{c}}\ll Tr H_{i}(x)\Phi H_{j}(y)\Phi^{+} \gg=\delta_{ij}(D^{H}(u)+D^{H}_{1}(u)+\textbf{u}^{2}\frac{\partial D^{H}_{1}(u)}{\partial \textbf{u}^{2}})-u_{i}u_{j}\frac{\partial^{2} D^{H}_{1}(u)}{\partial u^{2}}, 
\end{eqnarray} 
where $u = x-y$ and $\Phi(x,y)=Pexp(\int^{x}_{y}{A_{\mu}dz^{\mu}})$.
At zero temperature  both string tensions ($\sigma^{E}=\sigma^{H}=\sigma$) coincide and $\sigma$ forms the basic np scale,
which defines all hadron masses and the QCD scale in general.   The values of $ \sigma^{E, H} $ can be also obtained from  calculations  on the lattice (see for example \cite{TVCM5}). \par
The correlators $D^{E}$ and $D^{E}_{1}$ produce both the scalar confining interaction $V_{D}(r)$ and
the vector-like interaction $V_{1}(r)$:
\begin{eqnarray}
V_{D}(r)=2c_{\alpha}\int^{r}_{0}{(r-\lambda)d\lambda}\int^{\infty}_{0}d\nu D^{E}(\lambda,\nu)=V^{lin}_{D}(r)+V^{sat}_{D}(r)\\
V_{1}(r)=c_{\alpha}\int^{r}_{0}{\lambda d\lambda}\int^{\infty}_{0}d\nu D^{E}_{1}(\lambda,\nu), c_{fund}=1,c_{adj}=\frac{9}{4}
\end{eqnarray}
				From $V_{D}(r)$ we extract  the purely linear form $V_{D}^{lin}(r)$, and  for   $V_{1}(r)$ we  separate out the one gluon exchange, $V^{oge}$, $V_{1}(r)=V^{sat}_{1}+V^{oge}$, while the rest parts, $V_{d}^{sat}$ and $V_{1}^{sat}$ are saturating at large r,  thus  for the total potential below $T_{c}$ one   obtain: 
\begin{eqnarray}
V(r,T<T_{c})=V_{D}^{lin}(r)+V^{sat}_{1}(r)+V^{oge}(r)+V_{D}^{sat}(r)
\end{eqnarray}  
				It is worth emphasizing  that at low temperatures $V_{D}^{sat}(r)$ and $V^{sat}_{1}$  compensate each other (for details, see  appendix of \cite{TVCM5}). But at temperatures above $T_{c}$, $V_{D}(r)$ vanishes.  As for $V_{1}^{sat}(r,T)$,
 this quantity defines   Polyakov loops ($L_{i}$), i.e:
\begin{eqnarray}
L_{i}= exp(-c_{i}  \frac{V_{1}(\infty ,T)}{2T}),c_{fund}=1,c_{adj}=\frac{9}{4},i=adj,f \label{eqL}
\end{eqnarray}
				The contribution of  Polyakov loops  alone   gives   a reasonable agreement with the lattice results \cite{TVCM4}.However for an accurate  description of data one needs the CMC ingredient which we introduce below following \cite{TVCM8}. 
The relationship between pressure, volume and free energy is given by:
$$P_{gl}V_{3}=-F_{0}(B)$$
For the gluon contribution we obtain:
\begin{eqnarray}
P_{gl}=2(N^{2}_{c}-1)\int^{\infty}_{0}\frac{ds}{s}\sum_{n\ne 0}G^{n}(s),
G^{n}(s)=\int{(Dz)^{\omega}_{on}e^{-K}\hat{tr}_{a}<W^{a}_{\Sigma}(C_{n})>} 
\end{eqnarray}
\begin{eqnarray}
K=\frac{1}{4}\int^{s}_{0} d\tau(\frac{dz^{\mu}}{d\tau})^{2},
(Dz)^{\omega}_{xy}=\lim_{N \to \infty} \prod\limits^{n}_{m=1}\frac{d^{4}\zeta(m)}{(4\pi\epsilon)^{2}}\sum_{n=0,\pm,...}\frac{d^{4}p}{(2\pi)^{4}}exp[ip_{\mu}(\sum_{m=1}^{N} \zeta(m)-(x-y)_{\mu}-n\beta\delta_{\mu 4})]
\end{eqnarray}
where we use Fock-Feynman-Schwinger (FFS) formalism with Schwinger proper time s \cite{TVCM8}. $W^{a}_{\Sigma}(C_{n})$ is the adjoint  Wilson loop
defined for the gluon path $C_{n}$, which has both temporal (i4) and spacial projections (ij), and $\hat{tr}_{a}$  is the normalized
adjoint trace.

 CE and CM fields strengths in $T > T_{c}$ region correlate very weakly
due go the gauge-invariant field correlator in adjoint representation  $<E_{i}(x)B_{k}(y)\Phi(x,y)> \approx 0$
(see \cite{TVCM4,TVCM8}) and therefore both CE and CM projections of the $tr_{a}W^{a}_{\Sigma}(C_{n})$ can be factorized as shown in \cite{TVCM5}
\be  <W^{a}_{\Sigma}(C_{n})>=L^{(n)}_{adj}(T)<W_{3}>, \ee 
 \begin{center}for   $L^{(n)}_{i} \approx L^{n}_{i}$ for $T \le 1$ GeV, \end{center}
One can integrate out  the $z_{4}$ part of the path integral $(Dz)^{\omega}_{on}=(Dz_{4})^{\omega}_{on}D^{3}z$, and write the result as
\be G^{(n)}(s)=G^{(n)}_{4}(s)G_{3}(s),G^{n}_{4}(s)=\int (Dz_{4})^{\omega}_{on}e^{-K}L^{(n)}_{adj}=\frac{1}{2\sqrt{4\pi s}}e^{-\frac{n^{2}}{4T^{2}s}}L^{(n)}_{adj} \ee

For the Polyakov loops, one can obtain \cite{TVCM1} :
\be
L^{n}_{adj}=exp(-\frac{9}{4}J^{E}_{n}),
J^{E}_{n}=\frac{n\beta}{2}\int^{n\beta}_{0}d\nu(1-\frac{\nu}{n\beta})\int^{\infty}_{0}\zeta d\zeta D^{E}_{1}(\sqrt{\zeta^{2}+\nu^{2}})
\ee

and finally
\begin{eqnarray}
P_{gl}=\frac{N^{2}_{c}-1}{\sqrt{4\pi}}\int^{\infty}_{0}\frac{ds}{s^{3/2}}G_{3}(s)\sum_{n=\pm 1, \pm 2,...}e^{-\frac{n^{2}}{4T^{2}s}}L^{n}_{adj} \label{eqg} \\
G_{3}(s)=\int (D^{3}z)_{xx}e^{-K_{3d}}<\hat{tr}_{a}W^{a}_{3}> \label{eq3d1}
\end{eqnarray}
In a similar way one can consider the quark contribution :
\begin{eqnarray}
P_{q}=2N_{c}\int_{0}^{\infty}\frac{ds}{s}e^{-m^{2}_{q}s}\sum^{\infty}_{n=1}(-1)^{n+1}[S_{n}(s)+S_{-n}(s)], \label{18}
S_{n}(s)=\frac{1}{N_{c}}\int{(Dz)^{\omega}_{on}e^{-K}\hat{tr}<W_{\sigma}(C_{n})>}  
\end{eqnarray}

and the pressure acquires the form:
\begin{eqnarray}
P_{q}=\frac{4N_{c}}{\sqrt{4\pi}}\int^{\infty}_{0}\frac{ds}{s^{3/2}}e^{-m^{2}_{q}s}S_{3}(s)\sum_{n= 1, 2,...}(-)^{n+1}e^{-\frac{n^{2}}{4T^{2}s}}L^{n}_{f} \label{eqf}\\
S_{3}(s)=\int (D^{3}z)_{xx}e^{-K_{3d}}<\hat{tr}_{f}W^{f}_{3}> \label{eq3d2}
\end{eqnarray}
The equations (\ref{18}),(\ref{eqg})
provide a general  expression for  the free energy $-F=P_{g}+P_{q}$ and we can find all thermodynamic quantities and their dependence on parameters of the media.\par
It is necessary to make two important remarks.
1)  
  from the factor $ exp(-\frac{n^{2}}{4T^{2}s})$   and also from the contribution of CMC  follows the suppression of high order terms in (\ref{eqg}),(\ref{eqf}).  2)  the contribution    of CMC  in Single Line Approximation  dictates the form of the propagator  $G_{3}(s)$ and defines   the screening masses  $M_{D}$ in it. The latter,as shown in  \cite{VCM1,TVCM5} is growing with T and defines all dynamics at large T.\par
\section{The speed of sound in SU(3) Yang-Mills theory}
We start with  the speed of sound in  the case  of  SU(3) gluodynamics.  The Lagrangian is
\begin{eqnarray}
\mathcal{L}=-\frac{1}{4}G^{a}_{\mu\nu}G^{a}_{\mu\nu} \label{eql}\\
G^{a}_{\mu\nu}=\partial_{\mu}A^{a}_{\nu}-\partial_{\nu}A^{a}_{\mu}+gf^{abc}A^{b}_{\mu}A^{c}_{\nu}
\end{eqnarray}
where $G^{a}_{\mu\nu}$-  is Non-Abelian  field strength, $a=1,..,N^{2}-1,\mu,\nu=1..4$  In this model there is a  confinement-deconfinement phase  transition, of the weak first order, from the phase of the glueball gas  to the gluon plasma as known from  lattice studies(see e.g  \cite{lattglu4}), and from the FCM analysis \cite{TVCM8}).  In addition to the gauge symmetry of the Lagrangian (\ref{eql}), there is also a scale symmetry of the Lagrangian on the classical level,
\begin{eqnarray}
x \to \lambda^{-1} x, A^{a}_{\mu}(x) \to \lambda A^{a}_{\mu}(x) 
\end{eqnarray}        
  As a consequence , the trace of the energy-momentum tensor must vanish  $<T_{\mu\mu}>=0$. 
 One might expect that in the thermodynamic description the equality $ E = 3P $ (where $ E $ is the system energy  and $ P $ is the system  pressure) holds. However, it is  well known that  inclusion  of quantum effects for non-Abelian fields leads to the appearance of a mass scale.
From the lattice calculations  we know that the scaling symmetry in  SU(3) Yang-Mills  is significantly violated, especially in the confinement-deconfinement transition area \cite{lattglu}. 
As  mentioned  in the introduction, the speed of sound is an excellent indicator for this violation, therefore we expect to obtain in our calculations  that the speed of sound is  different from $1/3$ in the vicinity of $T_{c}. $  \par 
The second important fact is related to the type of transition. In \cite{opt}, it was suggested that in  case of pure  Yang-Mills the type of confinement-deconfinement transition depends on the number of colours. In  case of $ N_{c}=2 $ there must be a second-order phase transition \cite{opt},while in the  case of $ N_{c}=3 $   it is of the first order  \cite{opt2},\cite{lattglu4}.  From  the expression \begin{eqnarray}
C^{2}_{s}=\frac{s}{\frac{\partial \epsilon}{\partial T}}=\frac{\frac{\partial P}{\partial T}}{\frac{\partial \epsilon}{\partial T }} \label{sos}
\end{eqnarray}
one can see that at the point $ T = T_{c} $  a possible discontinuity  in the speed of sound  that is confirmed by calculations on the  lattice \cite{lattglu}.  \par 
The introduction of  FCM  for  QCD was described in the previous section and will be used below for numerical calculations.  In addition one can provide a qualitative analysis of resulting equations . One can use eq.(\ref{eqg}) and obtain the energy density: 
 
 \begin{eqnarray}
\epsilon+P=T\frac{\partial P}{\partial T}
\end{eqnarray}

 Writing $P=T^{4}f(T)$ one has:
\begin{eqnarray}
C^{2}_{s}=\frac{\frac{\partial P}{\partial T}}{T\frac{\partial^{2}P }{\partial T^{2}}}=\frac{1}{3}\frac{f+\frac{1}{4}Tf^{'}(T)}{f+\frac{2}{3}Tf^{'}(T)+\frac{1}{12}T^{2}f^{''}(T)}\approx \frac{1}{3}(1-\frac{5}{12}\frac{Tf^{'}(T)}{f(T)}+O(T^{2}f^{''}))
\end{eqnarray}
To understand the behavior  of  $f(T)$  we use  \cite{TVCM2} for $G_{3}(s)$
\begin{eqnarray}
P_{gl}=\frac{2(N^{2}_{c}-1)}{16\pi^{2}}\sum^{\infty}_{n=1}L^{n}_{adj}\int^{\infty}_{0}\frac{ds}{s^{3}}e^{-\frac{n^{2}}{4T^{2}s}}\phi(M^{2}_{0}s) \label{form1}
\end{eqnarray}
where $\phi=(\frac{M^{2}_{0}s}{sh(M^{2}_{0}s)})^{\gamma}$, $\gamma=1$ for oscillator form of colour magnetic confinement, and  $\gamma=1/2$ for the linear confinement. The screening mass $M_{0}$ is expressed via spacial string tension,  $M^{2}_{0}=a\sigma_{s}$, $a=8\gamma$,  and   $\sigma_{s}$ was obtained from the lattice data \cite{lattglu}:
\begin{eqnarray}
\sigma_{s}(T)=c^{2}_{\sigma}g^{4}(T)T^{2}, c_{\sigma}=0.566\pm 0.013
\end{eqnarray} 
  Changing variables in the integral ($\ref{form1}$) one  obtains:
\begin{eqnarray}
I_{n}(\kappa^{2})=\int^{\infty}_{0}\frac{du}{u^{3}}e^{-\frac{n^{2}}{4u}}\phi(u\kappa^{2}), \kappa=\frac{M^{2}_{0}}{T^{2}}
\end{eqnarray} 
$f(T)$ can be written as:
\begin{eqnarray}
f(T)=\frac{2(N^{2}_{c}-1)}{16\pi^{2}}\sum_{n=1}^{\infty}L_{adj}^{n}I_{n}(\kappa^{2})
\end{eqnarray}
The derivative $f^{'}(T)$ consists of  two terms $\frac{\partial L^{n}_{adj}}{\partial T}$ and $\frac{\partial I_{n}(\kappa^{2})}{\partial T}=I^{'}_{n}(\kappa^{2})\frac{\partial \kappa^{2}}{\partial T },\frac{\partial \kappa^{2}}{\partial T}=8\gamma c^{2}_{\sigma}\frac{\partial g^{4}}{\partial T},$
where:
\begin{eqnarray}
I^{'}_{n}(\kappa^{2})=\int^{\infty}_{0}\frac{du}{u^{3}}e^{-\frac{n^{2}}{4u}}\frac{\partial \phi}{\partial \kappa^{2}}\frac{\partial \kappa^{2}}{\partial T}
\end{eqnarray}   
and one obtains  that both $\frac{\partial \phi}{\partial \kappa^{2}}$ and $\frac{\partial \kappa^{2}}{\partial T}$  are negative,  so that 
 $\frac{\partial I}{\partial T}>0$. The same conclusion follows for $\frac{\partial L_{i}}{\partial T}$,
\begin{eqnarray}
\frac{\partial L^{n}_{adj}}{\partial T}=\frac{\partial}{\partial T}(e^{\frac{-9nV_{1}(\infty,T)}{8T}})>0
\end{eqnarray}
where $V_{1}(\infty,T)$ decreases with T for $T>T_{c}$, while $L^{n}_{adj}$ grows with T. Hence  one can deduce  that:
$f^{'}(T)>0$, for $T>T_{c}$ and consequently:

\begin{eqnarray}
C^{2}_{s}-\frac{1}{3} \approx -\frac{5}{36}\frac{f^{'}(T)}{f(T)}<0 \label{eq33}
\end{eqnarray}
Note, that the largest contribution  to $f^{'}(T)$ comes from $\frac{\partial L_{adj}^{n}}{\partial T}$ \par
 In the confinement area, $T<T_{c}$ we can also make some predictions. For the pressure of glueball resonans gas one can write:
 \be f_{gb} =\sum_i \frac{g_i}{2\pi^2} \sum^\infty_{n=1} \frac{m^2_i(T)}{n^2T^2} K_2 \left( \frac{nm_i}{T} \right) \equiv \sum_i \frac{g_i}{2\pi^2} \sum^\infty_{n=1} \frac{1}{n^4} k^{(i)}_n \label{27}\ee
 \be k^{(i)}_n (T) \equiv \frac{n^2 m^2_i(T)}{T^2} K_2 \left( \frac{nm_i}{T} \right).\label{28}\ee
 
 In (\ref{27}), (\ref{28}) one can take  into account, that confinement and string tension $\sigma (T)$ are $T$ dependent \cite{TVCM5}, so that
 \be m_i (T) =a(T) m_i(0), ~~ a(T) = \sqrt{\frac{\sigma_E (T)}{\sigma_E(0)}}.\label{29}\ee
  Note, that $k_n^{(i)} (T\to \infty) \to \frac12$, while for small $T, T\ll m_i(T)$ one has 
 \be k_n^{(i)}  (T\to 0) = \sqrt{\frac{\pi}{2}} \left( \frac{nm_i(0)}{T} \right)^{3/2} \exp \left( -\frac{nm_i (0)}{T} \right)\label{30}\ee
As a consequence for $T\ll m_i(0)$ the main  contribution comes from the  term with the lowest mass, and thus from (\ref{28}) :
	\be
	C^{2}_{s}=\frac{T}{m_{0}},T\to 0 \label{eq38}
	\ee

\begin{figure}[h!]
\begin{center}
\includegraphics[scale=0.3,width=17 cm]{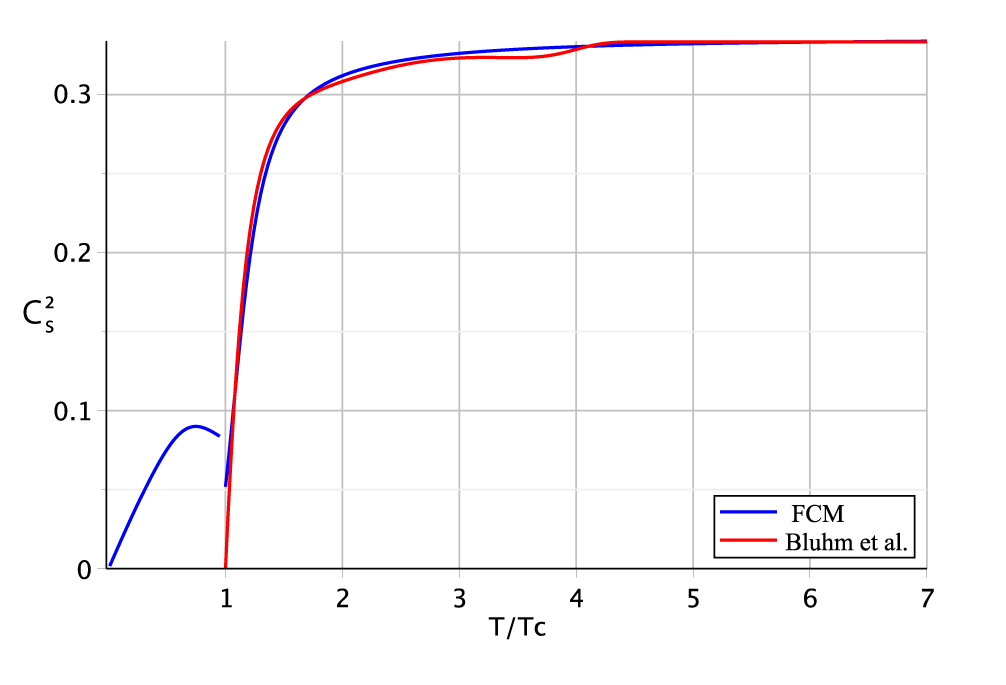}
\caption{The speed of sound  in FCM from (\ref{sos}) for SU(3)  in comparison with  lattice data \cite{fit} } \label{FIG1}
\end{center}
\end{figure} 
\par
\begin{figure}[h!]

           {\includegraphics[scale=0.35]{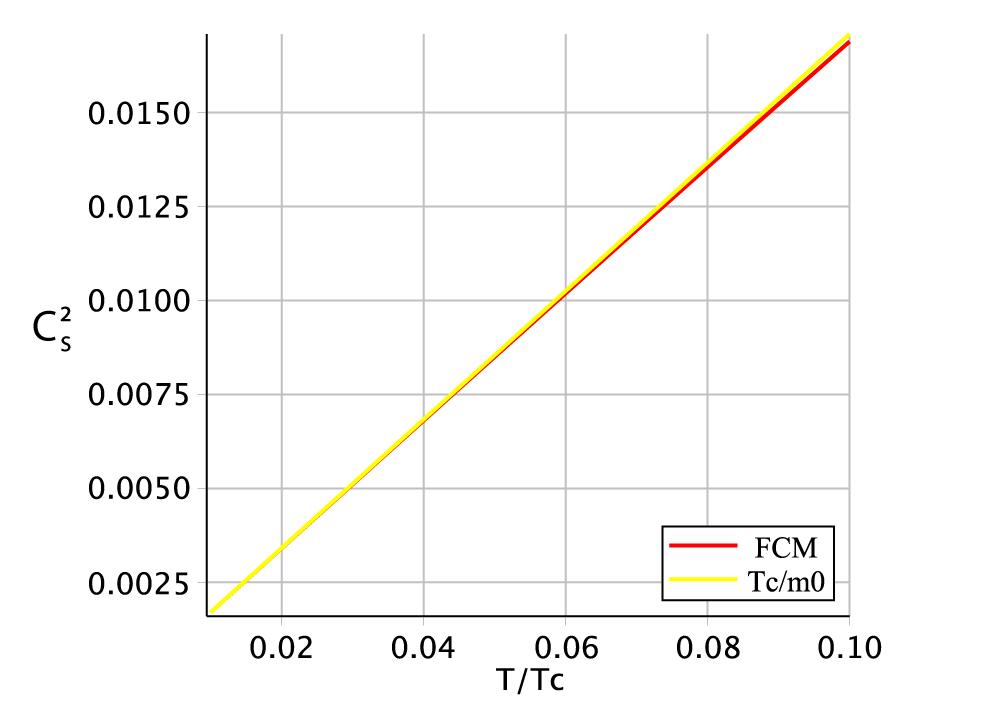}} 
									\caption{The speed of sound $C^{2}_{s}$ in FCM from (\ref{sos})  for SU(3) in the limit $T\to 0$ in comparison with (\ref{eq38}) } \label{FIG2}
		\end{figure}
As  one can see in FIG.1 there is  a good agreement between the predictions of the theory eqs.(\ref{eqg}),(\ref{sos})  and the lattice results\cite{lattglu}. We also obtain discontinuity at $T=T_{c}$, as it should be in case of the first  order phase transition. 
\newpage
 \section{The speed of sound in   QCD  with (2+1) flavours}

Here we shall consider  the speed of sound in case of QCD with $n_{f}=2+1$. We take $m_{u}, m_{d}$  equal to zero and  $m_{s}=0.1$ GeV.\par
As in the case of the Yang-Mills theory, some qualitative predictions  about  the value of the speed of sound can be made. First of all, because the confinement-deconfinement phase transition in this case is a crossover \cite{thLatt3},  the speed of sound can have a finite value and a local    minimum   in the crossover region. The second prediction is connected with the domain of extremely large T, where $C_{s}$ should be   close to $1/\sqrt{3}$,  because in this region   the scale anomaly is small relative to the typical energy scale  \cite{latt3} \par 
 
Similarly to the way it was done for $P_{g}$ in the last chapter, we can consider $P_{q}$,eq(\ref{eqf}) for the ensemble of quarks. We  use the pressure  from \cite{TVCM7}.
 
 \be P_q = \sum_f P^{(f)}_q, ~~ P^{(f)}_q = \frac{4N_c}{(4\pi)^2} \int \frac{ds}{s^3} e^{- {m^2_f s} } \sum_{n=1,2} (-)^{n+1} e^{-\frac{n^2}{4T^2 s}}   \cosh \left( \frac{\mu n}{T} \right) L_f^{(n)} \varphi_f(s)\label{34}\ee
 
 \be \varphi_f (s) = \left(\frac{M^2_f s}{\sinh  (M^2_f s)}\right)^\gamma, ~~ M^2_f = \frac49 M^2_0 = \frac49 a\sigma_s\label{35}\ee
 
 Similarly to (\ref{form1}), one obtains for $f_f(T)$ at $\mu=0$
 
 \be f_f (T) = \frac{4N_c}{(4\pi)^2}  \sum^\infty_{n=1} L_f^{(n)} I_n^{(f)} (\kappa^2) e^{-\frac{m^2_f}{T^2} u}\label{36}\ee
 with \be I_n^{(f)} = \int^\infty_0 \frac{du}{u^3} e^{-\frac{n^2}{4u}} \varphi_f (u\kappa^2_f), ~~ \kappa^2_f = \frac{M^2_f}{T^2}.\label{37}\ee
 
 Writing now $C^{2}_{s}-\frac{1}{3}=-\Delta_{qg}$ ,with  $\Delta_{qg}(T)= \frac{5}{36} T(\frac{f'_g(T)}{f}+ \frac{f'_f(T)}{f})$ and neglecting the relatively small T  derivatives of $\kappa^2_{f,g} = \frac{M^2_{f,g}}{T^2}$, one obtains for small $ \frac{m_{f}}{T} $
 \be  \Delta_{qg}\cong \frac{5}{36} \left\{ \sum_f \sum_n \frac{4 N_c}{(4\pi)^2}    e^{-\frac{m^2_f}{T^2} u}I_n^{(f)} \left(T\frac{\partial L_f^{(n)}}{\partial T}\right) + \sum_n \frac{2( N_c^2-1)}{(4\pi)^2}   I_n (\kappa^2) \left(T\frac{\partial L_f^{(n)}}{\partial T}\right)\right\}
 \label{38}\ee
 
 One can see that always $\Delta_{qg}>0$ and at large $T$, $T>250$ MeV, $\Delta_{qg}$ is small and tends to zero.
  In a similar way as for the glueball gas, one can treat the hadron resonance  gas of  mesons and baryons.Neglecting interaction between hadrons,one arrives at the small T limit (\ref{eq38}) for the sound velocity of the HRG with the pion mass for $m_{0}$   \par 
We compare our results with \cite{latt3} and \cite{thLatt2} in FIG.3.

   \begin{figure}[h!]
    \begin{center}
    \includegraphics[scale=0.35]{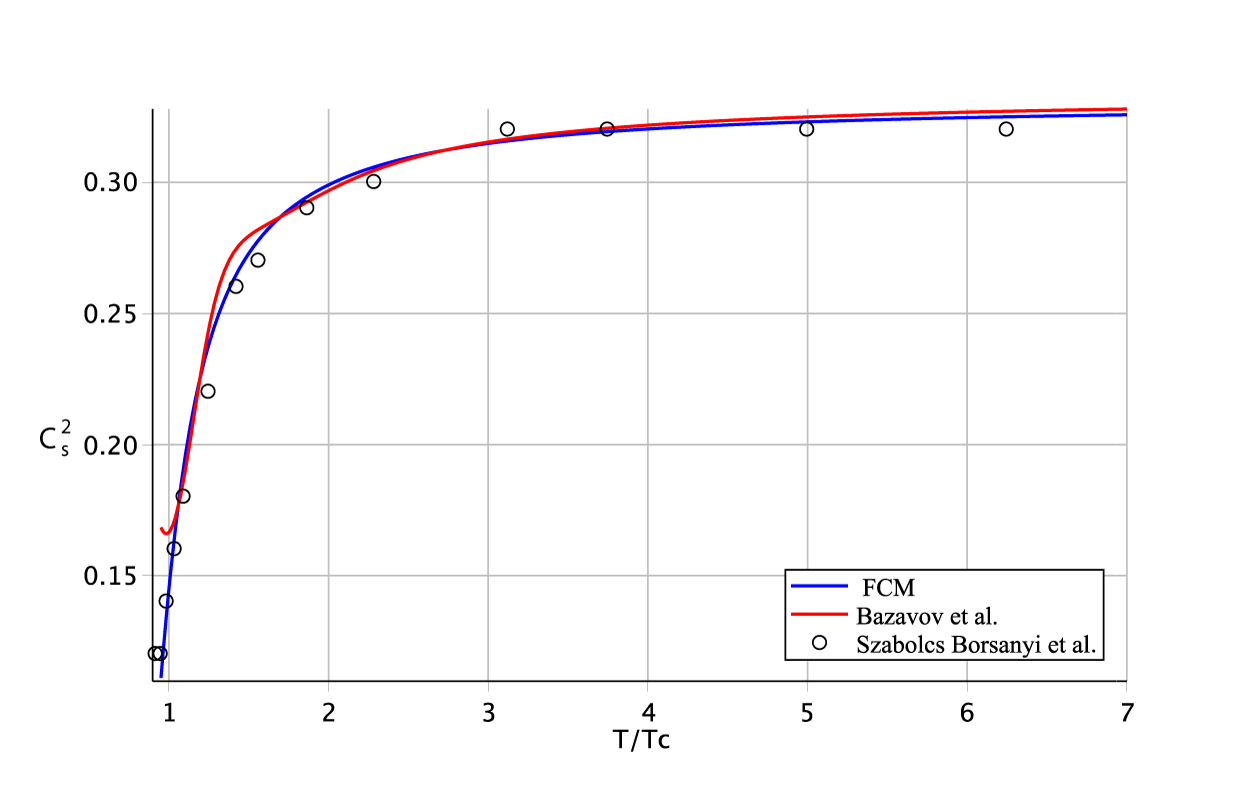}
		\caption{ The speed of sound in QGP from FCM, eqs.(\ref{sos},\ref{eqg},\ref{eqf}), in comparison with lattice data of Borsanyi et al  \cite{thLatt2}  and Bazavov et al \cite{latt3} }
		\end{center}
\end{figure}

\section{Numerical results and discussions}
To understand the behavior of $C_{s}^{2}$ as a function of T and compare it with the our qualitative predictions (\ref{eq33}) and (\ref{38}), as well as with lattice calculations, we have computed $P_{gl}(T)$ Eq. (\ref{form1}) with $\gamma=1/2$ and obtain $C^{2}_{s}$, using the definition (\ref{sos}) The results for $C^{2}_{s}$ are plotted in FIG.1 in comparison with the lattice data from \cite{fit}. One can see a good agreement with lattice curve for $C_{s}^{2}$ and a discontinuity  at $T=T_{c}$, as should be for the first order SU(3) transition. Both lattice and our results for $C_{s}^{2}$ are in the region $C_{s}^{2}\le 1/3$, which supports our qualitative conclusion in (\ref{eq33}).
It is interesting to study the behavior of $C_{s}^{2}$ at $T \to 0$, which was done in Fig.2, where the limiting relation (\ref{eq38}) is compared with our numerical data, showing a perfect  agreement. For the 2+1 QCD our numerical data are presented in Fig.3 in comparison with the lattice data of \cite{thLatt2} and \cite{latt3}. One can see again a good agreement of all results. Comparing with our analytic  predictions in (\ref{38}), $C^{2}_{s}-\frac{1}{3}=-\Delta_{qg}$, one finds that indeed $\Delta_{qg}$ is positive in the whole region $T>T_{c}$, and it is small  for $T>250$ MeV.\par
Thus  for the  QGP  one reveals the behavior, of the squared sound velocity, which never exceeds $1/3$ and can be called normal. However , already  at nonzero $\mu$ one might meet  with a new phenomenon, since $\mu$ enters $P_{q}$, as in (\ref{34}) via cosh$\frac{\mu n}{T}$, and this provides a negative sign of $\Delta_{qg}$ for large enough $\mu/T$.
One can expect also a strong deviation of $C^{2}_{s}$ in the presence of external  magnetic fields.Both this effects require additional studies and will be subject of further  publications. \par
The authors are grateful for useful discussions to M.A. Andreichikov and  B.O.Kerbikov, M.A.Zubkov. \par
 This work was done in the frame of the scientific project, supported by the Russian Foundation grant number 16-12-10414.

\end{document}